\documentclass[%
 superscriptaddress,
twocolumn,
 amsmath,amssymb,
 aps,
]{revtex4}

\usepackage{graphicx}
\usepackage{dcolumn}
\usepackage{bm}

\begin{document}

\preprint{APS/123-QED}

\title{Elastic ripening and inhibition of liquid-liquid phase separation}


\author{Kathryn A. Rosowski}
 \thanks{KAR and TS contributed equally}
 \affiliation{Department of Materials, ETH Z\"{u}rich, 8093 Z\"{u}rich, Switzerland.}
 
\author{Tianqi Sai}
 \thanks{KAR and TS contributed equally}
\affiliation{Department of Materials, ETH Z\"{u}rich, 8093 Z\"{u}rich, Switzerland.}

\author{Estefania Vidal-Henriquez}
\affiliation{Max Planck Institute for Dynamics and Self-Organization, 37077, G\"{o}ttingen, Germany}

\author{David Zwicker}
\affiliation{Max Planck Institute for Dynamics and Self-Organization, 37077, G\"{o}ttingen, Germany}

\author{Robert W. Style}
 \email{robert.style@mat.ethz.ch}
 \affiliation{Department of Materials, ETH Z\"{u}rich, 8093 Z\"{u}rich, Switzerland.}
 
\author{Eric R. Dufresne}
 \email{eric.dufresne@mat.ethz.ch}
 \affiliation{Department of Materials, ETH Z\"{u}rich, 8093 Z\"{u}rich, Switzerland.}

\date{\today}

\begin{abstract}

Phase separation has recently emerged as an important organizational principle in the dense and  heterogeneous environment within living cells. 
Here, we use a synthetic system to show that compressive stresses in a polymer network suppress phase separation of the solvent that swells it.  These stresses create a barrier to droplet nucleation that leads to robust, stabilized mixtures well beyond the liquid-liquid phase separation boundary. Network stresses not only alter the stability  of mixtures, but they also have a dramatic effect on the ripening of droplets.  Gradients in network stresses can drive a new form of ripening, where solute is transported down stiffness gradients.  This elastic ripening can be much faster than conventional surface tension driven Ostwald ripening.
\end{abstract}

\maketitle

Phase separation has recently emerged as an important route to compartmentalization within living cells.
It is widely assumed to be governed by the physics of liquid-liquid phase separation \cite{bran09,hyma14,pate15,shin17}.
However, the cell interior is not a simple liquid, but a crowded macromolecular stew including polymer networks that impart significant elasticity \cite{janm07}. 

Recent experiments have shown that network elasticity can dramatically impact liquid-liquid phase separation in swollen synthetic polymer networks.
In a homogeneous network,  droplets grow to a fixed size, controlled by the network stiffness \cite{styl18}.  
When the network has an anisotropic state of stress, droplets grow with scale-independent ellipsoidal shapes \cite{kim2018}.

In the nucleus of living cells, artificial phase separating domains formed preferentially in chromatin-poor regions \cite{shin18}.
After drops were triggered to grow in chromatin-rich regions, they  migrated toward chromatin-poor regions. 
Theory and simulation suggest that these observations could be driven by gradients in network stiffness \cite{styl18,shin18}. 
However, these exciting observations are not sufficient to establish the role of network elasticity in droplet nucleation and migration.  
There are two main challenges.
First, the heterogeneous mechanical properties of the nucleus remain unquantified.  Second, the nucleoplasm is a multicomponent mixture, and the chemical solubility of one component (\emph{i.e.} chromatin) can dramatically effect the solubility of other components (see 
\emph{e.g.} the phase diagrams of three-component mixtures \cite{veatch2003,vitale2003}).

 Here, we reveal the impact of network mechanics on droplet nucleation and ripening with experiments in a synthetic polymer system, where mechanical properties and chemical solubility can be tuned independently \cite{styl18}.
 We find that network elasticity can  fully suppress droplet nucleation deep inside the thermodynamically immiscible region, thus mechanically stabilizing  supersaturated mixtures. 
 The long-term stability of droplets is strongly affected by network elasticity.  
 While we observe no ripening in homogenous networks, in a mechanically heterogeneous network, solute moves from stiff to soft regions by diffusive transport through the dilute phase.
Superficially resembling surface tension driven Ostwald ripening, this \emph{elastic ripening} can be much faster and is driven by gradients of network elasticity.

\begin{figure}
\includegraphics[width=8cm]{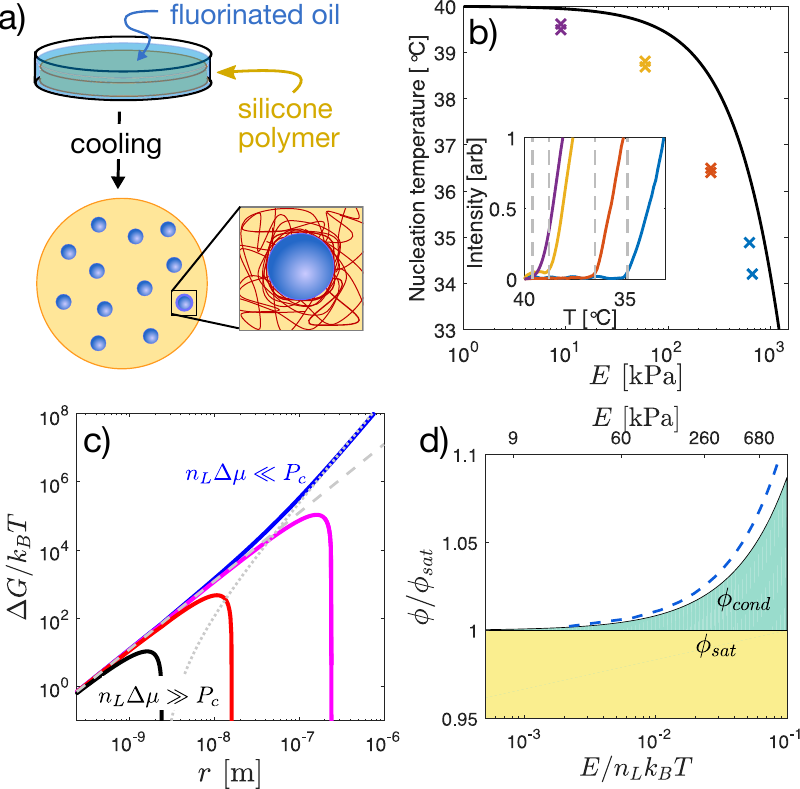}
\caption{\label{fig:nucleation} \emph{Network stiffness controls droplet nucleation.} (a) Schematic diagram of the method for nucleating and growing droplets in a polymer network.  (b) Measured nucleation temperature for samples saturated at $T_{sat}=40^\circ\mathrm{C}$ and cooled down at  $0.2^\circ$C /min, as a function of Young's modulus, $E$.  Inset shows the intensity signal used to identify the nucleation temperature. (c) Free energy landscape of a growing droplet, as given by Eq. \ref{eqn:nuc}.  Here, $E$ is set at 300 kPa, $r_m \approx2.4$ nm, $\gamma = 4 \mathrm{mN/m}$ \cite{kim2018} and the driving pressure $n_L \Delta \mu$ is varied from 50 kPa (blue line) to 5 MPa (black line).  Contributions from surface tension and elasticity are shown as dashed and dotted lines, respectively. (d) Fixed-temperature phase diagram showing the stability of the mixture with concentration and network stiffness. Mixtures are stable in the yellow region, unstable in the white region, and stable against droplet growth but unstable to demixing at the boundaries in the green region.  The solid black curve for $\phi_{cond}$ is given by Eq.\ref{eq:phicond}, assuming elastic cavitation and dilute solution approximation for $\Delta \mu$. The dashed blue line shows a non-dilute version of the theory with $\phi_{sat}$= 0.036, as described in the Supplement.}
\end{figure}

To investigate the impact of network elasticity on nucleation,  we drove phase separation in silicone gels (Gelest). First, we saturated them  in a bath of fluorinated oil (Fluorinert FC-770) at  $T_{sat}=40^\circ\mathrm{C}$ \cite{styl18}.  After several hours, the volume fraction of oil  saturated at $\phi_{sat} \approx 0.036$, independent of Young's modulus, $E$, as shown in Figures S3,S4. 
Then, we quenched samples in a temperature controlled microscope stage (TSA12Gi, Instec), while recording bright-field images with a 20X NA 0.5 objective. 
With sufficient undercooling, droplets nucleate and grow, shown schematically in Figure \ref{fig:nucleation}a.
Nucleation events were identified by quantifying the average pixel-wise intensity difference at each temperature from a reference image, which grows rapidly at the point of nucleation, as shown in the inset panel of Figure \ref{fig:nucleation}b. 
At quench rates below $1$ $^\circ\mathrm{C}/\mathrm{min}$, the nucleation temperature, $T_{nuc}$, has no significant dependence on the quench rate (Fig. S6).
Thus we used a quench rate of $0.2$ $^\circ\mathrm{C}/\mathrm{min}$ to measure the droplet nucleation temperature as a function of network stiffness from 9 kPa to 680 kPa.
The nucleation temperature depends strongly on the network stiffness. 
In flexible networks, $E=9$ kPa, droplets nucleated at only $0.5^\circ\mathrm{C}$ below  $T_{sat}$.
In stiff networks, $E=680$ kPa, droplets did not appear until the sample was cooled about $6^\circ\mathrm{C}$ below $T_{sat}$ (Fig. \ref{fig:nucleation}b).

These results suggest that network elasticity may constrain droplet nucleation.
When the radius of the droplet exceeds the network mesh size, $r_m \approx (k_BT/E)^{1/3}$ \cite{rubi03}, the network can squeeze the growing droplet.
However, as long as the pressure within a droplet exceeds a critical value, $P_C$, it can overcome elastic stresses and continue to grow  \cite{gent91,zimb07,kim2018}.
We assume here that $P_C=5E/6$, a classic result for nonlinear-elastic solids \cite{gent91}, which is a good approximation for many polymer networks \cite{zimb07}.  
This mechanical resistance is readily incorporated into classical nucleation theory, which considers the free energy, $\Delta G$, of a  droplet as a function of its radius, $r$, 
\begin{equation}
    \Delta G = -\frac{4 \pi}{3}r^3n_L \Delta \mu + 4 \pi r^2\gamma + \frac{4 \pi}{3}\Theta(r-r_m)(r^3-r_m^3)P_c. 
    \label{eqn:nuc}
\end{equation}
Here, as in simple condensation, nucleation is driven by the chemical potential difference between the dilute and condensed phases, $\Delta \mu$,  and resisted by interfacial energy, $\gamma$.
The number density of molecules in the condensed phase is $n_L$, and $\Theta(r)$ is the Heaviside step function.
The product $n_L \Delta \mu$ is equivalent to the stall pressure, $P_{st}$,  the maximum pressure a growing droplet can exert against its surroundings  \cite{styl18}.
The last term is a simple approximation of the mechanical work required to expand the cavity beyond the scale of the network mesh, \cite{gent91}.
It can have a dramatic impact on the free energy landscape governing droplet growth. 
Figure \ref{fig:nucleation}c shows the predicted size dependence of the droplet free energy for four different supersaturation levels in a 300 kPa sample. 
At the highest supersaturations ($n_L \Delta \mu=$ $5 ~\mathrm{MPa}$, black curve), elasticity has no impact on the free energy landscape and droplets can nucleate and grow normally.
At lowest supersaturation ($n_L \Delta \mu=$ $50~ \mathrm{kPa}$, blue curve), the droplet free energy monotonically increases with radius, completely forbidding droplet growth.

\begin{figure*}
\includegraphics[width=18cm]{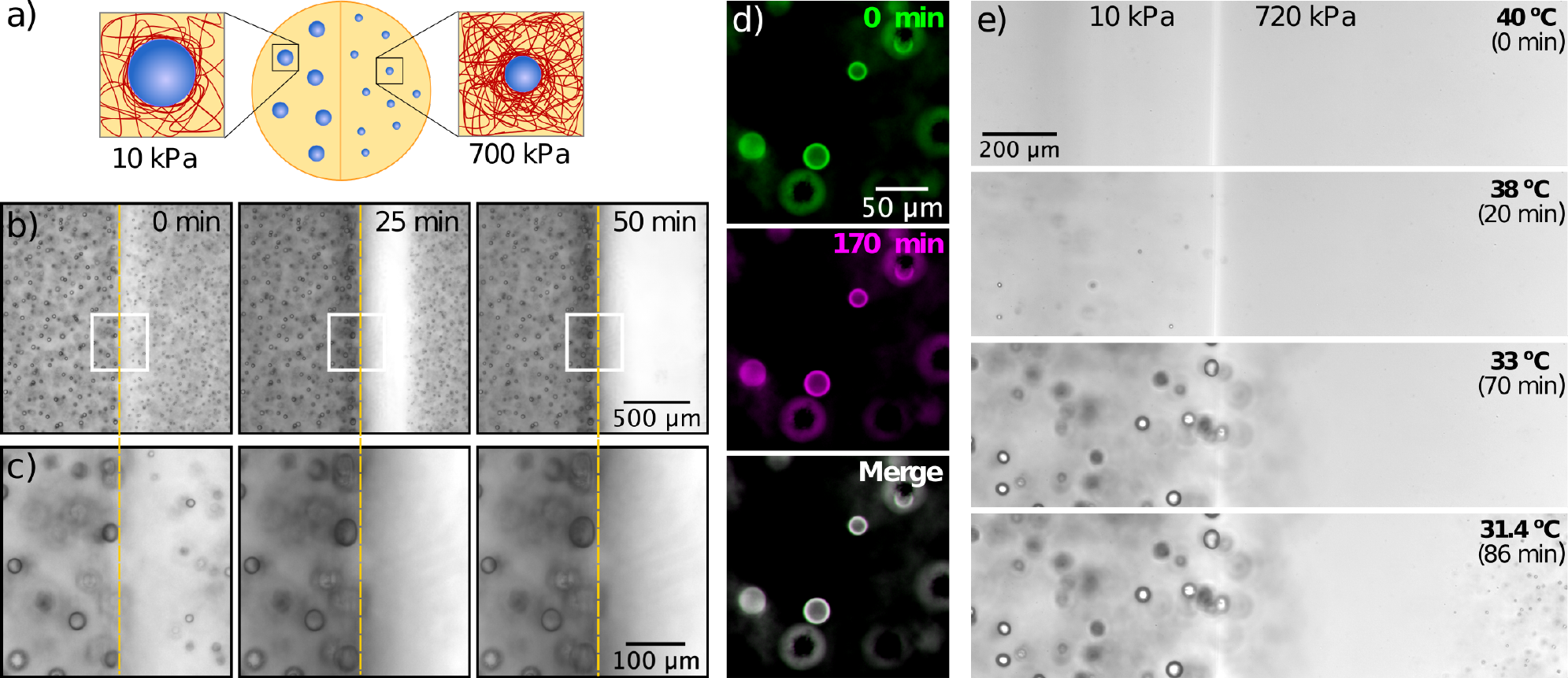}
\caption{\label{fig:gradient} \emph{Stiffness gradients drive solute transport and ripening.} (a) Schematic showing liquid droplets generated in a stiffness gradient.  (b)-(c)  Time-lapse bright-field images (10X NA 0.3) showing dissolution of droplets on the stiff side ($E=700$kPa) and growth of droplets on the soft side ($E=10$kPa). White boxes in (b) are shown at higher magnification in (c). (d) False color images of bi-disperse droplets grown in a homogeneous network with $E=80$ kPa. Green(magenta) panels show the sample at the 0(170) minute time point. The merged image shows almost complete overlay of the two channels (seen as white) (e) Side-by-side gradient samples of stiff ($E=720$kPa) and soft ($E=10$kPa)  cooled down at a  rate of $0.1^\circ$C /min. Orange dashed lines in (b),(c) indicate the interface between stiff and soft sides.}
\end{figure*}

Droplet nucleation is thus allowed only when $n_L \Delta \mu > P_c$.
Assuming the ideal form of the chemical potential of the dilute phase, $\Delta \mu= k_BT \ln \left( \phi/\phi_{sat} \right )$, we identify a minimal volume fraction of solute needed for condensation, \begin{equation}
  \phi_{cond} =  \phi_{sat} e^{P_C/n_L k_BT} \approx \phi_{sat} e^{5 E/ 6n_L k_BT}.
 \label{eq:phicond}
\end{equation} 
The ratio $E/n_L k_BT$ thus plays  a central role.
As it increases beyond one, 
the supersaturation required for droplet nucleation diverges exponentially.
In these experiments,  $n_L k_BT = 11~\mathrm{MPa}$, and $\phi_{cond}/\phi_{sat}$ reaches about 1.05, as shown in Figure \ref{fig:nucleation}d. 

We identify three regimes of stability.
When $\phi < \phi_{sat}$ (shaded yellow in Figure 1d), the mixture is stable, independent of elasticity.  When $\phi  > \phi_{cond}$ (shaded white), the mixture is unstable.
When $\phi_{sat} < \phi  < \phi_{cond}$ (shaded green) the supersaturated system is stable against nucleation and growth of droplets, but can demix by expelling solute beyond the boundaries of the polymer network.  
  In this regime, compositions that would be unstable in the absence of elasticity are completely stable in the limit of infinite system size (\emph{i.e} the thermodynamic limit).

To compare these predictions to our nucleation data, we determined $T_{cond}$, the temperature where Eq. \ref{eq:phicond} is satisfied, \emph{i.e.} $\phi_{sat}(T_{sat}) = \phi_{sat}(T_{cond}) \exp(P_C/n_Lk_BT_{cond})$.
Using independently measured values of 
 $\phi_{sat}(T)$ (Fig. S4), this yields the solid line  in Figure \ref{fig:nucleation}b. 
As expected, the stiffness dependence of $T_{cond}$ follows the measured trend of $T_{nuc}$, but at a consistently higher temperature, 
likely reflecting kinetic limitations to nucleation.
To further test our hypothesis that the supersaturated state is stable to nucleation of droplets, we cooled a 700 kPa sample to $T=38^\circ\mathrm{C}$, where $T_{cond}<T<T_{sat}$.  
At this undercooling, softer samples nucleate and grow droplets in a matter of seconds.  
However, we fixed the temperature at this value for $>60$ minutes, and found no evidence of nucleation in the stiffer sample.

Thus, elastic forces hinder nucleation and stabilize the supersaturated mixed state.
Similarly, we expect the dilute phase to remain supersaturated even after droplet growth is completed.
Consider a system quenched to $\phi>\phi_{cond}$, driving droplet nucleation.
As droplets grow, they deplete  solute from the dilute phase, and  $n_L \Delta \mu$ decreases.
When  $n_L \Delta \mu = P_c$,  the free energy liberated by condensation is just sufficient to overcome the mechanical work needed to deform the polymer network, and droplet growth is arrested.  
At this point,  $\phi=\phi_{cond}$, and the concentration in the dilute phase is stably supersaturated.

Since $\phi_{cond}$ increases with network stiffness, stiffness gradients should result in gradients of supersaturation, and transport of solute from stiff to soft regions \cite{webe17}.  This simple physical picture rationalizes transport down stiffness gradients observed in simulations \cite{shin18}.

To test this hypothesis, we nucleated and grew droplets in samples with stiffness gradients, shown schematically in Figure \ref{fig:gradient}a.  
Samples were saturated with oil at $T_{sat}=40^\circ \mathrm{C}$ to establish a uniform concentration $\phi_{sat}(T_{sat})$, and quenched quickly so that both sides nucleated and grew droplets.
Experimental data with a side-by-side step change in stiffness from 10 kPa to 700 kPa are shown in Figure \ref{fig:gradient}b,c.
As expected \cite{styl18}, we found larger droplets on the soft side than the stiff side.  
While droplets far from the interface were stable (Fig. S9 and  Movies S1,S2), droplets on the stiff side, adjacent to the interface, started to dissolve soon after growth.
Over time, more droplets dissolved on the stiff side, leaving a growing band of clear gel adjacent to the interface (Fig. \ref{fig:gradient}b, Movie S3).
Simultaneously, the droplets on the soft side, immediately adjacent to the interface, grew (Fig. \ref{fig:gradient}c). 
This experiment reveals significant solute transport down the stiffness gradient, supporting our hypothesis that the solute is still supersaturated when droplet growth is arrested.

This phenomenon resembles conventional surface tension driven Ostwald ripening \cite{dege04}, as small droplets shrink to feed the growth of larger droplets.
To rule this out, we performed two experiments.   First, we  tested if Ostwald ripening occurs between droplets in a homogeneous silicone network.  We produced a bi-dispersed droplet size distribution in a homogeneous 80kPa sample by making a sudden change in the quench rate during droplet growth, described in the Supplement. 
This method allowed us to produce droplets with radii of about $12$ and $9.5 ~\mathrm{\mu m}$, with a typical spacing of about $90 ~\mathrm{\mu m}$.  We monitored the size of the droplets for 170 minutes after the end of growth, and found no change in size (Fig. \ref{fig:gradient}d, Movie S4).  
Compare this to the 140 kPa gradient sample in Figure S10, with droplets of  $12$ and $10 ~\mathrm{\mu m}$ juxtaposed across the interface.  
In this case, smaller droplets more than $100~\mathrm{\mu m}$ from the larger ones are dissolved completely within 100 minutes.

Second, we repeated the experiment with a side-by-side stiffness gradient, as in Figure \ref{fig:gradient} b, but with slow and controlled cooling, at $0.1^\circ\mathrm{C}/\mathrm{min}$.  
As expected from our data on nucleation in homogeneous samples (Fig. \ref{fig:nucleation}), droplets appeared at higher temperatures on the soft side than on the stiff side  (Fig. \ref{fig:gradient}e).  
Additionally, droplets never appeared near the stiff side of the interface. 
This indicates that stiffness-driven gradients of supersaturation depleted solute from the stiff side  even before droplets could form there.    While this is consistent with our proposed mechanism, it is not consistent with surface tension driven Ostwald ripening, which requires droplets of different sizes. 

\begin{figure}
\includegraphics[width=8cm]{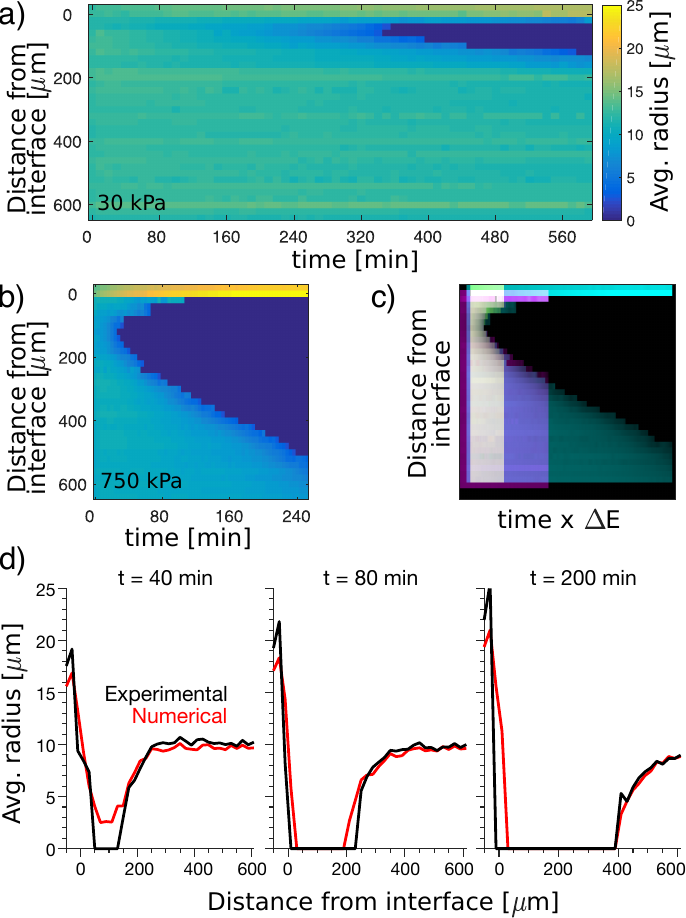}
\caption{\label{fig:kymograph_new} \emph{Rate of ripening increases with stiffness difference.} (a),(b) Kymographs showing average radius over time of droplets at different distances from the soft-stiff interface. The `soft' sides (z $<0\mathrm{\mu m}$) have $E=7$ kPa.  `Stiff' sides (z $>0\mathrm{\mu m}$) vary from $E=30$ kPa in (a) to $E=750$ kPa in (b) (further data Fig S10). (c) Superimposed kymographs with stiff sides of  $E=140$kPa (yellow), $E=330$kPa (magenta), and $E=750$kPa (cyan), where the time-scale is multiplied by the elastic modulus difference across the interface, and shifted so that the starting points are aligned. (d) Numerical simulation of 750kPa sample, shown with experimental data (see Movie S5 for further time points). Here, the simulation is based on simple diffusion in the dilute phase with the internal droplet pressure set by the stiffness of the gel.}
\end{figure}

Having ruled out surface tension as a driving force for ripening, we now quantify the effect of stiffness gradients.
We measured the average droplet size as a function of  time and distance from the soft-stiff interface, visualized as a kymograph in Figure \ref{fig:kymograph_new}a,b. 
Keeping the stiffness at $z<0$ fixed at 7 kPa, we varied the stiffness for $z>0$ from 30 to 750 kPa.
Results for the extremes are shown in Figure \ref{fig:kymograph_new}a,b, the rest in the Supplement.
In all cases, droplets on the stiff side disappear near the interface after a delay, and a dissolution front invades the stiff side. 
The dissolution front appears sooner, and moves faster, in stronger stiffness gradients.  
In Figure \ref{fig:kymograph_new}c, we stretch these kymographs by an amount proportional to the stiffness difference, $\Delta E$, and superimpose them.  Intriguingly, with  a small shift in starting time and starting position,  the shapes of the dissolution fronts match (Fig. \ref{fig:kymograph_new}c).
The rate of advance of the dissolution front therefore increases with $\Delta E$.

A simple model of solute transport captures the essential features of these results.
It is based on simple diffusion in the dilute phase, where the concentration is pinned to $\phi_{cond}\approx\phi_{sat}\exp(P_C/n_L k_\mathrm{B}T)$ at the surface of the droplets (see Supplement).
This simple theory, shows  quantitative agreement for large $\Delta E$, as shown in Fig. \ref{fig:kymograph_new}d (and Movie S5).
It overestimates the delay before front formation for smaller stiffness differences, as shown in Fig. S10.

In general, both elasticity and surface tension can  drive ripening. 
In Ostwald ripening, the driving pressure is given by the difference in Laplace pressure.  
In elastic ripening, the driving pressure emerges from  differences in pressure applied to the droplets by the polymer network, of order $E$.
The relative strength of the driving forces for Ostwald and elastic ripening are therefore captured by the elastocapillary number, $\gamma/ER$, \cite{styl15}.  
When this is much smaller than one, elastic forces dominate.
In these experiments, $\gamma / ER \approx \mathcal{O}(10^{-3})$.
But could elastic ripening occur in living cells?
For membraneless organelles, reported values of  $\gamma$ vary from $10^{-7}-10^{-4}$ N/m \cite{bran09, bran11b, feri16, tayl16}, and $R$ from $1 - 10~\mathrm{\mu m}$ \cite{bran09, feri16}.
The cytoskeleton can readily reach stresses on the order of kPa \cite{wang02, rots00, gard06}.  
Thus, the elastocapillary number of membraneless organelles could readily reach $\mathcal{O}(10^{-3})$, where elastic ripening dominates.

We have shown that a host polymer network can dramatically alter the nucleation and ripening of phase-separated liquid droplets. 
In classic liquid-liquid phase separation, these processes are controlled by interfacial energy.
Here, the structure of the polymer network introduces a new term into the free energy, reflecting that droplets can grow only when their internal pressure exceeds a critical pressure, of order Young's modulus.  
This splits the conventional phase boundary in two. 
One curve, $\phi_{sat}$, determines the saturation equilibrium of a system in contact with a reservoir of solute.  Another, $\phi_{cond}$, determines the limit of stability  of the mixture against droplet nucleation and growth.

These phenomena further suggest that living cells could regulate the localization of membraneless organelles through gradients of their mechanical properties. 
Since cells are well-known to regulate their heterogeneous mechanical properties \cite{hoff2006,janm07}, this physical hypothesis is biologically reasonable.
By contrast, recent physical modelling of the localization of phase-separated domains has focused on the role of activity gradients, \cite{bran09,zwic14,zwic17, webe19}.
We should note that the current work is limited to static networks with relatively simple rheology.
New phenomena are anticipated in living cells where the timescale of phase separation can be comparable to the viscoelastic relaxation timescale \cite{tana00} or the remodelling timescale of active networks \cite{blan14,need17}.

We acknowledge the SNF National Centre of Competence in
Research `Bio-Inspired Materials' for funding, as well as Larry Wilen, Sanat Kumar, and Tal Cohen for helpful discussions. 

The datasets generated and analysed during the current study are available from the corresponding authors on reasonable request.

\end{document}